\documentclass[12pt]{article}
\usepackage{amsmath,amssymb,amsthm}
\usepackage{geometry}
\begin{document}
\title{New Two Dimensional Massless Field Theory from Bagger-Lambert-Gustavsson Model}
\author{
  M. A. Santos\thanks{email: masantos@cce.ufes.br}
	\thanks{On leave from Universidade Federal do Esp\'{\i}rito Santo (UFES).}\\
	\em Centro Brasileiro de Pesquisas F\'isicas\\
	\em R. Dr. Xavier Sigaude 150, 22290-180 Rio de Janeiro - RJ, Brazil\\
	and\\
	I. V. Vancea\thanks{email: ionvancea@ufrrj.br}\\ 
	\em Grupo de F{\'i}sica Te\'orica e Matem\'atica F\'isica,\\ 
	\em Departamento de F\'{\i}sica, \\
	\em Universidade Federal Rural do Rio de Janeiro (UFRRJ),\\
	\em Cx. Postal 23851, BR 465 Km 7, 23890-000 Serop\'{e}dica - RJ,
	Brazil
}

\date{31 October 2008}

\maketitle

\begin{abstract}
By compactifying the Bagger-Lambert-Gustavsson model on $\mathbb{R}^{1,1} \times S^1$, we obtain a new two dimensional massless field theory with dynamical fields valued in the Lie 3-algebra $\mathcal{A}$ coupled with an $SO(1,1)$ scalar and vector field which are valued in the set $End(\mathcal{A})$ of the endomorphisms of the Lie 3-algebra. In the limit $g_{BLG} \rightarrow \infty$ the theory reduces to a supersymmetric Lie 3-valued generalization of the Green-Schwarz superstring in the light-cone gauge.    
\end{abstract}

\newpage

\section{Introduction}

The lifting of the Nahm equation describing the fuzzy geometry of the D1-D3 intersection to the Basu-Harvey equations for M2-M5 system \cite{Basu:2004ed} has drawn the attention to the gauge theories with Lie $3$-algebras $\mathcal{A}$. The first $d=2+1$ $N=8$ superconformal field theory in which the Basu-Harvey equation plays the role of the BPS equation was recently proposed by Bagger and Lambert and Gustavsson (BLG) in \cite{Bagger:2006sk,Bagger:2007jr,Bagger:2007vi,Gustavsson:2007vu,Gustavsson:2008dy} and it was conjectured to be an effective field theory on the world-volume of a stack of M2-branes. That interpretation is correct in particular systems of M2-branes such as M-branes on M-folds \cite{Mukhi:2008ux,Lambert:2008et} while in other cases the theory should be modified. Nevertheless, the original BLG-model represents an important field theory by itself as well as due to the insights it provides in the analysis of the degrees of freedom of the M-theory.   

Since its discovery, important advances in understanding the structure of the BLG-model and its relation with the Lie $n$-algebras have been made. By analysing the representations of the $SO(4)$ gauge symmetry, it was shown that the $OSp(8|4)$ superconformal symmetry is parity conserving and it was conjectured that the BLG-theory would be unique \cite{Bandres:2008vf,VanRaamsdonk:2008ft}. In \cite{Gauntlett:2008uf}, the same conclusion was drawn from the compatibility between the positive definite metric and the triple product. The boundary terms in the BLG-theory and the supersymmetry preserving marginal and mass deformations were discussed in \cite{Berman:2008be,Hosomichi:2008qk,Song:2008bi,Ahn:2008ya,Krishnan:2008zm}. In \cite{Jeon:2008bx}, the authors discussed and classified the BPS states in the BLG-theory. The scalar field responsible for the reduction of the multiple M2-branes to the multiple D2-branes was interpreted in \cite{Banerjee:2008pd}. The gauge invariance of the theory and its ghost structure were investigated in \cite{Bandres:2008kj,Gomis:2008be}. Extensions of the theory to Lie 3-algebra with Lorentz metric were discussed in  \cite{Gomis:2008uv,Benvenuti:2008bt,Ho:2008ei,Ezhuthachan:2008ch,Cecotti:2008qs}.  Algebraic properties and different models based on BLG-theory were discussed extensively in the literature \cite{Papadopoulos:2008sk,Papadopoulos:2008gh,Morozov:2008cb,Gran:2008vi,Ho:2008nn,Shimada:2008xy,Morozov:2008rc,Honma:2008un,Fuji:2008yj,Li:2008ez,
Hosomichi:2008jd,FigueroaO'Farrill:2008zm,Park:2008qe,Passerini:2008qt,Mauri:2008ai,Bergshoeff:2008ix,deMedeiros:2008bf,Blau:2008bm,Furuuchi:2008ki,jff,Sochichiu:2008jm,Bandos:2008fr,Bonelli:2008kh}. Properties of the partition functions were recently approached in \cite{Hanany:2008qc,Bedford:2008hn}. The supersymmetric construction of supersymmetric field theories for condensates of M2-branes was given in \cite{Bandos:2008jv,Cederwall:2008vd,Cherkis:2008qr}. A non-linear extension of the BLG-model has been proposed in \cite{Iengo:2008cq}.

One of the most interesting problems of the BLG-model is its relation with two dimensional field theories which arise naturally in the string theory. Important progress in this direction was made in \cite{Mukhi:2008ux,Li:2008ya} where it was shown that the $d=1+2$ maximally supersymmetric Yang-Mills on a stack of N D2-branes in $D = 10$ dimensional spacetime can be obtained from the BLG-model by a Higgs mechanism which generates Yang-Mills dynamics from topological Chern-Simons terms. However, the M2-branes can be dimensionally reduced to F1-string. Also, it is known that the M2-M5 system is obtained as the strong coupling, $D=11$ limit of the F1-D4 configuration. That suggests that one should study the dimensional reduction of the BLG-model to two dimensions. 

In this work we are going to investigate the compactification of the BLG-theory on $\mathbb{R}^{1,1} \times S^1$ \footnote{The problem of compactification of the three dimensional space of extended BLG-models was addressed in \cite{Gomis:2008cv,Lin:2008qp}. For the compactification of the world-volume in the Nambu-bracket realization of the Lie 3-algebra see \cite{Bandos:2008df}.}. The theory contains the $\mathcal{A}$ valued dynamical fields $X^{I}$ and $\Psi^{A}$ from the vector and spinor representation of $SO(8)$, respectively, as well as the topological fields $A_{\alpha}$ and $\Phi$ which are an $SO(1,1)$ vector and scalar, respectively, and take values in $End(\mathcal{A})$. The massless compactification modes produce a new type of field theory in two dimensions which has the Lie 3-algebra structure. 
In general, the terms containing $A_{\alpha}$ and $\Phi$ break the two dimensional conformal symmetry, but it is not clear to us whether some of the original $N=8$ superconformal symmetry is still preserved. If the conformal invariance in two dimensions is imposed, these terms vanish leaving behind a supersymmetric two dimensional field theory for $X^{I}$ and $\Psi^{A}$ which represents a simple generalization of the Green-Schwarz superstring in the light-cone gauge to a Lie 3-algebra valued superstring. This interpretation holds in the limit $g_{BLG} \rightarrow \infty$ if $\mathbb{R}^{1,1} \times S^1$ is embedded in to $M^{1,9} \times S^1$ with the compactified direction of spacetime indentified with the compactified direction of the world-volume of the M2-branes. If a transversal direction of the spacetime is compactified instead, the $SO(8)$ symmetry is broken to $U(1) \times SO(7)$ structure. 

This paper is organized as follows. In Section 2, we are going to review the results from the BLG-theory. The two dimensional action is obtained in Section 3. In the last section, we discuss the weak/strong coupling limit of the two dimensional theory as an independent non-associative field theory in two dimensions and in the context of M-theory.     

\section{BLG-theory}

In this section we are going to review the BLG-theory following \cite{Bagger:2006sk,Bagger:2007jr,Bagger:2007vi,Gustavsson:2007vu}. The model describes an effective field theory living on the world-volume of a stack of M2-branes and it is formulated in terms of a set of an Euclidean Lie 3-algebra valued massless fields $\{ X^{I}(x), \Psi(x), A_{\mu}(x) \}$. Here, the indices of the bosonic fields $I = 1, 2, \ldots ,8$ and $\mu = 0,1,2$ refer to the vectorial representations of the transversal and longitudinal groups $SO(8)$ and $SO(1,2)$, respectively. The field $\Psi(x)$ is a $d=11$ Majorana spinor subjected to the $SO(8)$ chirality constraint 
\begin{equation}
\left( \Gamma^{012} \right)^{A}_{B} \Psi(x)^{B} = -\Psi(x)^A ,
\label{chirality}
\end{equation}
where $A, B$ are spinor indices of eleven-dimensional Majorana spinor. If $\{ T^{a} \}_{a = 1,2,\ldots N}$, denotes the (linear space) basis of the underlying $N$-dimensional Lie 3-algebra $\mathcal{A}$, the fields have the following decomposition
\begin{align}
X^{I}(x) = X^{I}_{a}(x) T^{a} ,
\nonumber\\
\Psi(x) = \Psi_{a}(x) T^{a} ,
\nonumber\\
A_{\mu}(x) = A_{\mu ab}(x) t^{ab},
\label{field-decomposition}
\end{align}
where $t^{ab} = \left[T^{a},T^{b},\cdot \right]$ is defined in terms of the triple product $\left[ \cdot, \cdot, \cdot \right]: \mathcal{A} \times \mathcal{A} \times \mathcal{A} \rightarrow \mathcal{A}$. (Our conventions for the Lie 3-algebra are the ones from \cite{Bagger:2007jr}.) If the dimension of the elements of $\mathcal{A}$ is neglected, the number of the bosonic degrees of freedom exceeds the number of the fermionic degrees of freedom by one. Therefore, in order to produce a supersymmetric theory, the Lagrangian should contain a topological term for $A_{\mu}$. The supersymmetric action proposed in \cite{Bagger:2006sk,Bagger:2007jr} has the following form
\begin{align}
S = \frac{1}{g^{2}_{BLG}}\int d^3 x 
  \left\{ 
      - \frac{1}{2} \left( D_{\mu}X^{aI}\right) \left(D^{\mu}X^{I}_{a} \right)
      + \frac{i}{2} \bar{\Psi}^{a} \Gamma^{\mu} D_{\mu} \Psi_{a} 
      + \frac{i}{4} \bar{\Psi}_{b} \Gamma_{IJ} X^{I}_{c} X^{J}_{d} \Psi_{a} f^{abcd}
      -  V(X)
  \right. 
\nonumber \\
  \left. 
      + \frac{1}{2} \epsilon^{\mu \nu \lambda} 
        \left( f^{abcd} A_{\mu ab} \partial_{\nu} A_{\lambda cd} 
               + \frac{2}{3} f^{cda}_{~ ~ ~ g} f^{efgb} A_{\mu ab} A_{\nu cd} A_{\lambda ef} 
        \right)
  \right\} .
\label{BL-action}
\end{align}
The covariant derivatives are defined by the following relations
\begin{align}
  D_{\mu } X^{I}_{a} = \partial_{\mu } X^{I}_{a} - \tilde{A}_{\mu ~ a}^{~ b} X^{I}_{b} , 
\label{covar-der-X}\\
  D_{\mu } \Psi_{a} = \partial_{\mu } \Psi_{a} - \tilde{A}_{\mu ~ a}^{~ b} \Psi_{b} ,
\label{Covar-der-Psi}
\end{align}
where $\tilde{A}_{\mu ~ b}^{~ a}  = f^{cda}_{~ ~ ~ b} A_{\mu cd}$. The potential $V(X)$ has the form
\begin{equation}
  V(X) = \frac{1}{12}f^{abcd}f^{efg}_{~ ~ ~ d} X^{I}_{a} X^{J}_{b} X^{K}_{c} X^{I}_{e} X^{J}_{f} X^{K}_{g}
       = \frac{1}{2} \mbox{Tr} \left( \left[ X^I, X^J, X^K \right],\left[ X^I, X^J, X^K \right] \right) .
\label{V-potential}
\end{equation}
The action (\ref{BL-action}) has the $SO(8)$ R-symmetry that acts on $X^{I}$. Also, it is invariant under the on-shell supersymmetry transformations
\begin{align}
\delta X^{I}_{a} & = i \bar{\epsilon} \Gamma^{I} \Psi_{a}
\label{susy-1},\\
\delta \Psi_{a} & = D_{\mu } X^{I}_{a} \Gamma^{\mu } \Gamma^{I} \epsilon 
                - \frac{1}{6} X^{I}_{b} X^{J}_{c} X^{K}_{d} f^{bcd}_{~ ~ ~ a} \Gamma^{IJK} \epsilon 
\label{susy-2},\\
\delta \tilde{A}_{\mu ~ a}^{~ b} & = i \bar{\epsilon} \Gamma_{\mu } \Gamma_{I} X^{I}_{c} \Psi_{d} f^{cdb}_{~ ~ ~ a} 
\label{susy-3},
\end{align}
where the spinorial parameter $\epsilon$ has opposite chirality from $\Psi$ for the unbroken supersymmetry, i. e.  $\Gamma^{012} \epsilon = \epsilon$. The equations of motion derived from (\ref{BL-action}) are
\begin{align}
  \Gamma^{\mu } D_{\mu } \Psi_{a} + \frac{1}{2} \Gamma_{IJ} X^{I}_{c} X^{J}_{d} \Psi_{b} f^{cdb}_{~ ~ ~ a} & = 0 ,
\label{eq-mot-1}\\
  D^{2} X^{I}_{a} - \frac{i}{2} \bar{\Psi}_{c} \Gamma^{I}_{J} X^{J}_{d} \Psi_{b} f^{cdb}_{~ ~ ~ a} - \frac{\partial V}{\partial X^{I a}} & = 0 ,
\label{eq-mot-2}\\
  \tilde{F}_{\mu \nu ~ a}^{~ ~ b} + \epsilon_{\mu \nu \lambda} \left( X^{J}_{c} D^{\lambda} X^{J}_{d} + 
  \frac{i}{2} \bar{\Psi}_{c} \Gamma^{\lambda} \Psi_{d} \right) f^{cdb}_{~ ~ ~ a} & = 0 ,
\label{eq-mot-3} 
\end{align}
where the gauge field is defined as 
\begin{equation}
  \tilde{F}_{\mu \nu ~ a}^{~ ~ b} = \partial_{\nu } \tilde{A}_{\mu ~ a}^{~ b} - \partial_{\mu } \tilde{A}_{\nu ~ a}^{~ b} 
                                  - \tilde{A}_{\mu ~ c}^{~ b} \tilde{A}_{\nu ~ a}^{~ c} 
                                  + \tilde{A}_{\nu ~ c}^{~ b} \tilde{A}_{\mu ~ a}^{~ c} .
\label{equation} 
\end{equation}
In order that the equations of motion be compatible with the gauge and supersymmetry transformations, the structure constants of $\mathcal{A}$ must satisfy the fundamental identity (see \cite{Bagger:2007jr,Gustavsson:2007vu}). The unitarity of the theory could be guaranteed, in principle, if the trace form metric of $\mathcal{A}$ is positive definite. There is just one no-trivial Lie 3-algebra $\mathcal{A}$ that satisfies these restrictions, namely $\mathcal{A}_4$ \cite{Gauntlett:2008uf,Ho:2008bn,Papadopoulos:2008sk}. The uniqueness of the algebraic structure of the effective field theory has been a strong motivation for generalizing the theory given in the relation (\ref{BL-action}) to algebras without a positive definite metric and to new Lie 3-algebras \cite{Song:2008bi,Morozov:2008cb,Gran:2008vi,Gomis:2008uv,Benvenuti:2008bt,Ho:2008ei}.

\section{BLG-Model on $\mathbb{R}^{1,1} \times S^1$}

In this section we are going to derive the two dimensional massless field theory obtained from the compactification of the BLG-model on $\mathbb{R}^{1,1} \times S^1$. The three dimensional metric is $g_{\mu\nu}= \mbox{diag}\left(  -,+,R^{2}\right)$. We denote by $y$ the compactified direction $x^{2}=y=R\theta$ where $ \theta \in \left[  0,2\pi \right] $. The fields are required to be periodic with respect to the direction $y$
\begin{align}
X_{a}^{I}\left(  x^{\alpha},y+L\right)  &= X_{a}^{I}\left(  x^{\alpha
},y\right)  ,
\label{per-X}\\
A_{\mu ab}\left(  x^{\alpha},y+L\right)  & = A_{\mu ab}\left(  x^{\alpha},y\right)  ,
\label{per-A}\\
\Psi_{a}^{A}\left(  x^{\alpha},y+L\right)  & = \Psi_{a}^{A}\left(  x^{\alpha },y\right)  ,
\label{per-Psi}
\end{align}
where $L=2\pi R $. As usual, the fields can be Fourier expanded in terms of the eigenfunctions $Y_{n}(y)$ of the operator $\square_y$ that satisfy the ortho-normalization relation
\begin{equation}
\frac{1}{L} \int \nolimits_{0}^{L} dy ~ Y^{\ast}_{n}(y) Y_{m}(y) = \delta_{m,n},
\label{orthonorm}
\end{equation}
where $m,n \in \mathbb{Z}$.
From the reality of $X_{a}^{I}, A_{\mu ab}$ and $\Psi_{a}^{A}$ it follows that their Fourier modes should satisfy the following relations
\begin{equation}
X_{\left( -n \right)}^{aI} \left(  x^{\alpha}\right) = X_{n}^{aI}\left(  x^{\alpha}\right) ~~,~~ 
A_{\alpha ab,\left( -n \right)} \left(  x^{\alpha}\right) = A_{\alpha ab, n} \left(  x^{\alpha}\right) ~~, ~~ 
\Psi_{\left(  - n \right)  }\left(  x^{\alpha}\right)  =\Psi_{\left(  n\right)}^{\ast}\left(  x^{\alpha}\right),
\label{reality-rel}
\end{equation}
where $\alpha = 0, 1$ is the two dimensional worldsheet index. The compactification of $X_{a}^{I}$ and $A_{\mu ab}$ is standard. In order to compactify the spinor field we introduce the dreibein $e_{\mu}^{~ i}$ in the usual way by the relation 
$g_{\mu\nu} = e_{\mu}^{~ i} e_{\nu}^{~ j}\eta_{ij}$ with the inverse $h_{i}^{~ \nu}$ defined by $e_{\mu}^{~ i} h_{i}^{~ \nu}=\delta_{\mu}^{~ \nu}$. Here, $i,j = 0, 1$ are two dimensional tangent space indices. Then the spinor action takes the form 
\begin{equation}
\frac{i}{2g_{BL}^{2}}\int d^{3}x \left[ R \overline{\Psi }^{aM}
\left(  \Gamma^{\alpha}\right)  _{MN}D_{\alpha}\Psi_{a}^{N} + 
\overline{\Psi}^{aM} \left(  \Gamma^{2} \right)_{MN} D_{2} \Psi_{a}^{N}\right]. 
\label{spinor-comp}
\end{equation}
The compactification modes have the following masses 
\begin{equation}
m_{X n} \propto n R^{-2} ~~,~~ m_{A n} \propto n R^{-1} ~~,~~ 
m_{\Psi n} \propto n R^{-1}. 
\label{masses}
\end{equation}
In order to obtain the massless field theory in $\mathbb{R}^{1,1}$, one should take the limit $R \rightarrow 0$ and truncate the Fourier expansion to the massless modes.  After some lengthy but straightforward
computations and after rescaling the fields by a factor of $\left(  2\pi R\right)  ^{1/2}$, one can write down the following action
\begin{align}
S_{0} &  =\frac{\pi R}{g^{2}}\int d^{2}x\left[  -D_{\alpha}X^{aI}D^{\alpha
}X_{a}^{I}-\frac{1}{R^{3}}\Phi_{cd}f^{cdba}X_{b}^{I}\Phi_{lp}f_{~~~a}%
^{lps}X_{s}^{I}\right.  
\nonumber\\
&  +f^{abcd}\epsilon^{\alpha\beta}\left(  A_{\alpha ab}\partial_{\beta}%
\Phi_{cd}+\Phi_{ab}\partial_{\alpha}A_{\beta cd}\right)  +i\left(
\overline{\Psi}^{a}\Gamma^{\alpha}D_{\alpha}\Psi_{a}-\frac{1}{R}\overline
{\Psi}^{a}\Gamma^{2}\Phi_{cd}f_{~~~a}^{cdb}\Psi_{b}\right)  
\nonumber\\
&  +\frac{i}{2}f^{abcd}\overline{\Psi}_{b}\Gamma_{IJ}X_{c}^{I}X_{d}^{J}%
\Psi_{a}-\frac{1}{6}f^{abcd}f_{~~~d}^{efg}X_{a}^{I}X_{b}^{J}X_{c}^{K}X_{e}%
^{I}X_{f}^{J}X_{g}^{K}
\nonumber\\
&  +\left.  \frac{2}{3}f_{~~~g}^{abc}f^{defg}\epsilon^{\alpha\beta}\left(
A_{\alpha ab}A_{\beta cd}\Phi_{ef}-A_{\alpha ab}\Phi_{cd}A_{\beta ef}%
+\Phi_{ab}A_{\alpha cd}A_{\beta ef}\right)  \right]  .\label{effect-action}%
\end{align}
Here, $g=g_{BLG}R^{-\frac{1}{2}}$ is the two dimensional coupling constant. We
have denoted by $X_{a}^{I}$, $A_{\alpha ab}$ and $\Psi_{a}$ the zero modes of
the corresponding three dimensional fields, and by $\Phi_{ab}$ the zero mode
of $A_{2ab}$. The notation with covariant derivative has been maintained for its simplicity, with the two dimensional covariant derivative given by the usual relation
\begin{equation}
D_{\alpha}=\partial_{\alpha}+A_{\alpha cd}f_{~~~a}^{cdb}%
.\label{two-dim-cov-deriv}%
\end{equation}

The action (\ref{effect-action}) describes a new two dimensional effective field theory. Its massless dynamical field content is given by $X^{I}$ and $\Psi$ in the vector and spinor representations of $SO(8)$, respectively. The fields $A_{\alpha}$ and $\Phi$, which are in the $SO(1,1)$ vector and scalar representations, respectively, are non dynamical. The theory displays explicitly the original Lie 3-algebra structure. A simple algebra leads to the following equation of motion for $X^{M}_{m}$
\begin{align}
&D_{\alpha} D^{\alpha}X^{mM}
-\frac{1}{2 R^{3}}\left( f^{cdma} f^{lps}_{~~~ a} X_{s}^{M}\Phi_{cd}\Phi_{lp}
 +f^{cdba} f^{lpm}_{~~~ a} X_{b}^{M}\Phi_{cd} \Phi_{lp}
\right)
\nonumber\\ 
&+ \frac{i}{2} f^{abmd}\overline{\Psi}_{b}%
\Gamma_{MJ}X_{d}^{J}\Psi_{a}
-\frac{1}{12}\sum\limits_{k=1}^{3}\sum\limits_{j=1}^{6}%
K^{a_{1}..a_{j}..a_{6}}X_{a_{1}}^{I_{1}}... \hat{X}_{a_{j}}^{I_{k}%
}...X_{a_{6}}^{I_{3}}\delta_{M}^{I_{k}}\delta_{a_{j}}^{m}=0,
\label{eq-mot-X}
\end{align}
where $K^{a_{1}..a_{j}..a_{6}}\equiv f^{a_{1}a_{2}a_{3}d}f_{~~~~~~d}^{a_{4}a_{5}a_{6}}$ and $\hat{~}$ denotes a missing term. The spinorial equations of motion have the following form
\begin{align}
& \left(  \Gamma^{\alpha}\right)_{B}^{C}%
D_{\alpha}\Psi_{c}^{B}
-\frac{1}{ R}
\left(\Gamma^{2}\right)_{B}^{C}\Phi_{ed}f^{edb}_{~~~ c}\Psi_{b}^{B}
+\frac{1}{2}f^{aed}_{~~~c}\left(  \Gamma_{IJ}\right)_{B}%
^{C}X_{e}^{I}X_{d}^{J}\Psi_{a}^{B}=0 ,
\label{eq-mot-spin-1}\\
& D_{\alpha} \overline{\Psi}_{cA}
\left(  \Gamma^{\alpha}\right)^{AC}  
+ \frac{1}{R}\overline{\Psi}_{A}^{a}
\left(  \Gamma^{2}\right)^{AC}\Phi_{ed}f^{ed}_{~~c a} 
-\frac{1}{2}%
f_{a}^{~bed}\overline{\Psi}_{bA}\left(  \Gamma_{IJ}\right)^{AC}X_{e}^{I}%
X_{d}^{J} = 0.
\label{eq-mot-spin-2}
\end{align}
By varying the action (\ref{effect-action}) with respect to $A_{\alpha ab}$  we obtain the following equation 
\begin{align}
&
f^{ijab}\epsilon^{\gamma\alpha}\partial_{\alpha}\Phi_{ab}
-\frac{i}{2}\overline{\Psi}^{a} \Gamma^{\gamma}%
f^{ijb}_{~~~ a}\Psi_{b}
+  f^{ijba} X_{b}^{I}\partial^{\gamma
}X_{a}^{I}-f^{ijba}X_{b}^{I}A_{rs}^{\gamma}f^{rst}_{~~~ a} X_{t}^{I}
\nonumber\\
&+\frac{1}{3}\left[ f^{ijc}_{~~~ g} f^{defg} 
\epsilon
^{\gamma\beta}A_{\beta cd}\Phi_{ef}+f_{~~~ g}^{abi}f^{jefg}\epsilon
^{\alpha\gamma}A_{\alpha ab}\Phi_{ef}-f_{~~~ g}^{ijc}f^{defg}\epsilon
^{\gamma\beta}\Phi_{cd}A_{\beta ef} \right.
\nonumber\\
&- \left. f_{~~~ g}^{abc}f^{dijg}%
\epsilon^{\alpha\gamma}A_{\alpha ab}\Phi_{cd}+f_{~~~ g}^{abi}f^{jefg}%
\epsilon^{\gamma\beta}\Phi_{ab}A_{\beta ef}+f_{~~~ g}^{abc}%
f^{dijg}\epsilon^{\alpha\gamma}\Phi_{ab}A_{\alpha cd} \right]  = 0.
\label{eq-mot-A}
\end{align}
Finally, the variation of the action with respect to $\Phi_{ab}$ leads to the following equation
\begin{align}
& f^{mncd}\epsilon^{\alpha\beta}\partial_{\alpha}A_{\beta cd}-  
\frac{i{\pi}^{1/2}}{ R^{1/2}}\overline{\Psi}^{a}\Gamma^{2}
\Phi_{cd}f_{~~~ a}^{mnb}\Psi_{b}
-\frac{1}{R^{3}} f^{mnba}X_{b}^{I}\Phi_{lp}f_{~~~ a}%
^{lps} X_{s}^{I}
\nonumber\\
&+\frac{1}{3}\left[  f_{~~~ g}^{abc}f^{dmng}%
\epsilon^{\alpha\beta}A_{\alpha ab}A_{\beta cd}-f_{~~~ g}^{abm}%
f^{nefg} \epsilon^{\alpha\beta}A_{\alpha ab}A_{\beta ef}+f_{~~~ g}%
^{mnc} 
f^{defg}\epsilon^{\alpha\beta}A_{\alpha cd}A_{\beta ef}\right]  = 0.
\label{eq-mot-Phi}
\end{align}
Recall that the fields $A_{\mu ab}$ are the components of a three dimensional topological vector field. Then one can see that $A_{\alpha ab}$ and $\Phi_{ab}$ are non dynamical fields in two dimensions since the equations of motion  (\ref{eq-mot-A}) and (\ref{eq-mot-Phi}) do not produce any dynamical term. Therefore, these two equations should be interpreted as algebraic constraints on the dynamics of the fields $X^{I}$ and $\Psi$.

\section{Discussion}

The main feature of the new type of two dimensional massless field theory obtained in the previous section is its non-associative structured inherited naturally from the BLG theory. In the compactification limit $R \rightarrow 0$, the action (\ref{effect-action}) represents a field theory on $\mathbb{R}^{1,1}$ with a Lie 3-algebra structure. It has the same action in the case of a positive as well as of a non-positive definite metric on $\mathcal{A}$. The dynamical fields in two dimensions are $X^I$ and $\Psi$, respectively, and they are valued in the Lie 3-algebra $\mathcal{A}$. They are coupled with a topological two dimensional vector field $A_{ab}$ and a scalar field $\Phi_{ab}$ which are anti-symmetric endomorphisms from $End(\mathcal{A})$. 

A first step to interpret this theory at $R \rightarrow 0$ is to understand the predominant terms in different limits of its coupling constant $g$ that is determined by the coupling constant $g_{BLG}$ of the BLG-theory. Let us suppose that the range of $g_{BLG}$ is the full positive semi-axis $\mathbb{R}^{+}$. Then the weak coupling limit of (\ref{effect-action}) can be obtained from the $g_{BLG} \rightarrow 0$ limit of the BLG theory. The predominant interaction terms in decreasing strength order are: $\Phi X \Phi X$; $V(X)$; $\overline{\Psi} \Phi \Psi$; $\overline{\Psi} X X \Psi , A X A X$; $ \overline{\Psi} A \Psi , A X \partial X , \left( AA \Phi - A \Phi A + \Phi A A \right) $. However, rescaling the fields by the $\left( 2 \pi R \right)^{1/2}$ factor which amounts to the rescaling of the coupling constant ${g'}^2 \sim g^2 / R$ shows that the weak coupling corresponds to $g' = \mbox{constant}$. Then one can see that the predominant interactions of the dynamical fields are with the scalar field $\Phi$, i. e. $\Phi X \Phi X $ and $\overline{\Psi} \Phi \Psi$. 

The strong coupling limit of the theory on $\mathbb{R}^{1,1}$ can be obtained either from $g_{BLG} = \mbox{constant}$ or $g_{BLG} \rightarrow \infty$. In the case of $g_{BLG} = \mbox{constant}$, the field $\Phi$ still interacts strongly with the $X$'s fields and much weaker with the $\Psi$ field (the latter is of the same magnitude as $g_{BLG}$.) This suggests that a possible supersymmetry of the system in $X$ and $\Psi$ could be broken in the strong coupling regime which is a quite puzzling situation because one would expect a superconformal theory at strong coupling, from the original $N = 8$ superconformal symmetry. (Note that the theory is not invariant under the two dimensional conformal transformations due to conformal breaking terms in the action.) On the other hand, in the strong coupling limit obtained from $g_{BLG} \rightarrow \infty$ the coupling between $X$ and $\Phi$ is undetermined and the interaction of both $X$ and $\Psi$ with the $\Phi$ field can have the same strength. Thus, the supersymmetry of the two dimensional theory could be preserved in the strong coupling limit of BLG model. 

At a fixed energy scale, the effective coupling constant $g^{2}_{eff}=g^2/E$ of the theory given by the action (\ref{effect-action}) depends on $g_{BLG}$, too. From that, one can see that the weak coupled BLG can produce a low energy field theory in $\mathbb{R}^{1,1}$ at $g^2 = \mbox{constant}$ which is either strongly coupled or undetermined, e. g. if $g^2 \sim E$ and the energy scale is small but finite. Another possibility is $g^2 \rightarrow 0$ for which the effective field theory is either undetermined or has a weak constant coupling if the energy scale is low but finite. However, the high energy theory is generally weakly coupled. The strong coupled BLG theory generates a strongly coupled low energy theory in two dimensions and an either weakly coupled or an undetermined theory at high energies. 

Let us turn now to another important property of theory given by (\ref{effect-action}) that can be noted by compactifying a spacetime direction along $x^1$ or $x^2$, and by taking the limit $g_{BLG} \rightarrow \infty$ while keeping $R$ constant. Since the fields $X^{I}$'s have only transversal components, they will stay in $\mathbf{8_{v}}$. If the two dimensional fields are supposed to be univalued, the relations (\ref{per-X})-(\ref{per-Psi}) continue to hold. As mentioned above, the invariance of the action (\ref{effect-action}) under the two dimensional conformal transformations should not hold in general. In fact, by passing to unscaled fields, the only surviving terms in this limit are the kinetic terms for $X$ and $\Psi$ fields and $f^{abcd}\epsilon^{\alpha\beta}\left(  A_{\alpha ab}\partial_{\beta} \Phi_{cd}+\Phi_{ab}\partial_{\alpha}A_{\beta cd}\right)$. It is easy to verify that if the invariance of the three terms under an arbitrary conformal transformation is imposed, the action (\ref{effect-action}) takes the familiar form
\begin{equation}
\frac{1}{g^2}\int d^{2}x \left( - \frac{1}{2}
\partial_{\alpha}X^{aI}\partial^{\alpha}X_{a}^{I} + \frac{i}{2} \overline{\Psi}^{a}\Gamma^{\alpha}\partial_{\alpha}\Psi_{a} \right).
\label{GS-action}
\end{equation}
One is tempted to interpret the equation (\ref{GS-action}) as a generalization of the GS superstring action in $D = 10$ light-cone gauge. The equation (\ref{GS-action}) would correspond actually to $N= \mbox{dim} \mathcal{A}$ 
Lie 3-algebra valued superstrings. 
If instead of compactifying the spacetime along the longitudinal direction $x^1$ or $x^2$ one compactifies along a transversal one, e. g. $X^{8}_{a}  \sim X^{8}_{a} + 2 \pi \rho n \xi^{8}_{a}$, where $\rho$ is the spacetime compactification radius, $n \in \mathbb{Z}$ and $\xi^{8}_{a}$ is a constant Lie 3-algebra value vector, the symmetry of the bosonic fields breaks to $U(1) \times SO(7)$. This case was analysed in \cite{Hanaki:2008cu} in the context of $N=6$ Chern-Simons theory where the correct Nahm equation for multiple D2 ending on D4 was obtained.

In the above analysis, the main hypothesis that $g_{BLG} \in \mathbb{R}^{+}$ holds only if the BLG-model is understood as an independent field theory in three dimensions. In the context of M-theory, the interpretation of the BLG-model as the particular case $N=2$ of the $SU(N) \times SU(N)$ Chern-Simons theory developed in \cite{Aharony:2008ug} suggests that the coupling constant $g_{BLG}$ have an upper nonzero bound. Indeed, from the quantum consistency of the action, the coupling constant should be quantized as $g_{BLG}^{-2} \in \mathbb{N}$, which would correspond to the $k$-th level of the $SU(2) \times SU(2)$ Chern-Simons theory. Then the interpretation of the two dimensional theory follows from the general case of a constant and finite coupling $g_{BLG}$ discussed in the previous paragraph. In the compactification limit $R \rightarrow 0$ the two dimensional theory is strongly coupled or undetermined for $k \rightarrow \infty$, while in the decompactification limit it is weakly coupled. The main implication of the finite maximum value of $g$ is that at low energies one obtains a strongly coupled effective field theory on $\mathbb{R}^{1,1}$ which probably is conformal invariant in the moduli space. 
However, if the BLG-theory is weakly coupled, the field theory on $\mathbb{R}^{1,1}$ has undetermined coupling constant.

To conclude, by simply performing the compactification of the BLG-theory on the $S^1$, we have obtained a non-associative field theory in two dimensions. If it is viewed as an independent field theory in two dimensions, (\ref{effect-action}) presents interesting behavior at strong and weak couplings. In particular, it provides a generalization of the Green-Schwarz superstring to a Lie 3-algebra valued superstring if the two dimensional conformal invariance is imposed.
However, when interpreted in the context of M-theory, the effective field theory from (\ref{effect-action}) is strongly coupled and the superstring generalization is lost. Nevertheless, the effective coupling constant is undetermined at high energies and the superstring interpretation could still hold. 

As was mentioned in the introduction, since the M2-M5 system is the strong coupling, $D = 11$ description of the F1-D4 intersection from Type IIA theory, it would be interesting to see if there is any relationship between the theory obtained in this paper and the effective field theory of F1-strings, perhaps from the point of view of the more general effective field theory on M2-branes given in \cite{Aharony:2008ug}. Also, that could be a good motivation for a deeper study of the properties of the present theory as an independent field theory in two dimensions and in the context of M-theory.


\section*{Acknowledgments}

M. A. S. would like to thank to J. A. Helay\"{e}l-Neto and A. M. O. de Almeida for hospitality at LAFEX-CBPF where this work was accomplished. I. V. V. would thank to J. A. Helay\"{e}l-Neto for discussions. The authors acknowledge the constructive correspondence with an anonymous referee that helped to correct and clarify the results and their interpretation.
The research of I. V. V. has been partially supported by FAPERJ Grant E-26/110.099/2008.


\end{document}